\documentclass[11pt,twoside]{article}

\usepackage{asp2006}
\usepackage{epsf}
\usepackage{psfig}
\usepackage{lscape}

\begin{document}
\title{SBF: multi-wavelength data and models}   
\author{M. Cantiello\altaffilmark{1,2}, G. Raimondo\altaffilmark{2}, J.P. Blakeslee\altaffilmark{1}, 
E. Brocato\altaffilmark{2}, M. Capaccioli\altaffilmark{3}}
\altaffiltext{1}{Dep. of Physics and Astronomy, Washington State University, Pullman, WA 99164}
\altaffiltext{2}{INAF-Oss. Astronomico di Teramo, Via M. Maggini, 64100, Teramo, Italy}    
\altaffiltext{3}{INAF-Oss. Astronomico di Capodimonte, Vicolo Moiariello 16, 80131, Napoli, Italy}

\begin{abstract}
Recent applications have proved that the Surface Brightness
Fluctuations (SBF) technique is a reliable distance indicator in a
wide range of distances, and a promising tool to analyze the physical
and chemical properties of unresolved stellar systems, in terms of
their metallicity and age. We present the preliminary results of a
project aimed at studying the evolutionary properties and distance of the
stellar populations in external galaxies based on the SBF method. 

On the observational side, we have succeeded in detecting I-band SBF
gradients in six bright ellipticals imaged with the ACS, for
these same objects we are now presenting also B-band SBF
data. These B-band data are the first fluctuations magnitude
measurements for galaxies beyond 10 Mpc.

To analyze the properties of stellar populations from the data,
accurate SBF models are essential. As a part of this project, we have
evaluated SBF magnitudes from Simple Stellar Population (SSP) models
specifically optimized for the purpose. A wide range of chemical
compositions and ages, as well as different choices of the photometric
system have been investigated. All models are available at the
Teramo-Stellar Populations Tools web site: www.oa-teramo.inaf.it/SPoT.

\end{abstract}

We have measured B- and I-band SBF magnitudes for 6 elliptical
galaxies observed with the ACS camera on board of HST: NGC\,1407,
NGC\,3258, NGC\,3268, NGC\,4696, NGC\,5322 and NGC\,5557. Concerning
I-band images, their high S/N ratio allowed us to obtain SBF
measurements in different regions of the galaxies -- 5 concentric
annuli (Cantiello et al. 2005). On the contrary, the B-band images
have low S/N ($\sim$1), and SBF amplitudes can be measured only in one
single annulus. The reliability of these B-band measurements has been
verified via numerical simulations, by using a procedure which is able
to reproduce realistic images of elliptical galaxies, including the
stellar SBF signal.


The general lack of B-band SBF data hampered up to now a detailed
comparison with models, our observational data represent the first
sample of B- and I-band SBF measurements for a fair sample of distant
galaxies.  Figure 1 (left panels) shows the comparison of absolute SBF
magnitudes versus (B-I)$_0$ color data with SSP models from the Teramo
Stellar Populations Tools group (SPoT models, Raimondo et
al. 2005). SBF and color data appear generally well reproduced by
means of standard SSP models in the $\bar{M_I}$ vs. (B-I)$_0$ panel.
However, there is a considerable mismatch between SBF models and data
for some objects in the $\bar{M_B}$ vs. (B-I)$_0$ panel. Such
disagreement does not depend on the distance modulus adopted to
estimate the absolute SBF magnitudes, in fact the same mismatch is
present also in the distance-free SBF-color vs. color (B-I)$_0$
(Cantiello et al. 2006, ApJ submitted). In addition, adopting other
standard SSP models from literature (e.g. Blakeslee et al. 2001) or
also non-standard SSP models (e.g. alpha enhanced models) the
disagreement is not removed.

\begin{figure}[!ht]
\plotone{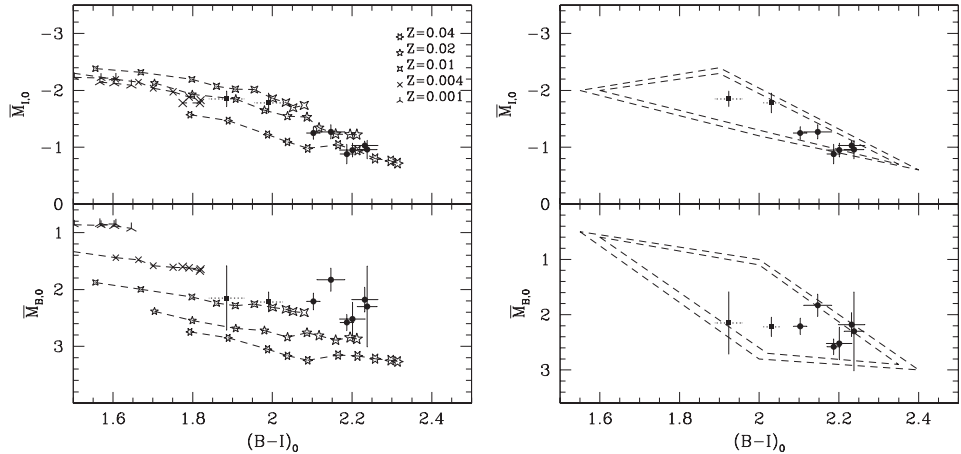}
\caption{Left panels: SBF absolute magnitudes vs. the (B-I)$_0$ color
derived from HST data (full dots). Full squares mark the only two
other galaxies with literature data. SSP models are from the SPoT
group for the labeled chemical compositions and 2 Gyr $\leq$ t $\leq$ 14 Gyr
(symbols of increasing size mark older ages). Right panels: same data
as left panels but compared to CSP models.}
\end{figure}
One possible solution seems to be the use of Composite Stellar
Populations (CSP). In Figure 1 (right panels) we compare the Blakeslee
et al. (2001) CSP models with the present data. These CSP models are
obtained combining SSP models in such a way to mimic, at least
approximately, the evolution of an elliptical galaxy. With these
models the disagreement between SBF data \& models disappears, as it
is completely accounted for by CSP with a fraction of old and
metal-poor (t$\sim$ 14 Gyr, [Fe/H]$\sim$-1.3) stars as high as 8\%,
combined with a dominant contribution from an old and metal rich
stellar component.


In conclusion, our data seem to show that while the integrated
properties of some galaxies might be well interpreted within the
scenario of classical SSP models, there are few objects whose
observational properties can only be interpreted by means of more
complex stellar populations systems. In this view, SBF and SBF colors,
coupled with classical photometric data appear to be a very
interesting tool to understand the properties of the unresolved
stellar systems in distant galaxies.


\end{document}